\documentclass[3p]{elsarticle}

    \makeatletter
    \def\ps@pprintTitle{%
      \let\@oddhead\@empty
      \let\@evenhead\@empty
      \let\@oddfoot\@empty
      \let\@evenfoot\@oddfoot
    }
    \makeatother

\usepackage{subfigure}      %\usepackage{wrapfig}		
\usepackage{float}					%\usepackage{fullpage}
\usepackage{placeins}	

\usepackage{amssymb,amsmath,amsfonts} 
\usepackage{graphicx}				\usepackage{color}
\usepackage{wrapfig}				\usepackage{subfigure}

\def\eeq{\relax}
\def\beq#1#2\eeq{\begin{equation}\label{#1}#2\end{equation}}
\def\bal#1#2\eal{\begin{align}\label{#1}#2\end{align}}
\def\bse#1#2\ese{\begin{subequations}\label{#1}#2\end{subequations}}
\def\ba{\begin{aligned}}   \def\ea{\end{aligned}}
\def\Xint#1{\mathchoice
{\XXint\displaystyle\textstyle{#1}}%
{\XXint\textstyle\scriptstyle{#1}}%
{\XXint\scriptstyle\scriptscriptstyle{#1}}%
{\XXint\scriptscriptstyle\scriptscriptstyle{#1}}%
\!\int}
\def\XXint#1#2#3{{\setbox0=\hbox{$#1{#2#3}{\int}$}
\vcenter{\hbox{$#2#3$}}\kern-.5\wd0}}

\def\dashint{\Xint-}

\def\dd{\operatorname{d}} 
\def\Im{\operatorname{Im}}
\def\Re{\operatorname{Re}}
\def\T{\operatorname{T}}

\def\IE{\operatorname{\Sigma}}
\begin{document} 
\title{Acoustic integrated extinction }  
\author{ Andrew N. Norris}
\address{Mechanical and Aerospace Engineering, Rutgers University, %\\
Piscataway, NJ 08854-8058, USA}

%\subject{mechanical engineering, structural engineering, mechanics}
%\keywords{acoustics, scattering, integrated extinction}
%\corres{Andrew Norris\\
%\email{norris@rutgers.edu}}				

%%%%%%%%%%%%%%%%%%%%%%%%%%%%%%%%%%%%%%%%%%%%%%%%%%%%%%%%%%%%%%%%%%%%%%%%%
%\def\singlespacing{\baselineskip=13pt}	\def\doublespacing{\baselineskip=18pt}
%\singlespacing%\doublespacing

%\pagestyle{myheadings}\markright{  \currfilename \qquad    ~~~~~~\today}
%

%\tableofcontents
\begin{abstract} %%%%%%%%%%%%%%%%

The integrated extinction (IE) is  defined as the integral of the  scattering cross-section as a function of wavelength.  Sohl et al.\ \cite{Sohl07} derived an IE expression for     acoustic  scattering that is causal,   i.e.\  the scattered wavefront  in the forward direction arrives later than the incident plane wave in the background medium.  The IE formula was based on  electromagnetic results, for which scattering is causal by default.  Here we derive a  formula  for the acoustic IE that is valid for causal and non-causal scattering.   The general result is expressed  as an integral of the time dependent forward scattering function.  The IE reduces to a finite integral for  scatterers with zero long-wavelength monopole and dipole amplitudes.    Implications for acoustic cloaking are discussed and a new metric is proposed for broadband acoustic transparency. 

\end{abstract}   %%%%%%%%%%%%%%%%
\maketitle  

%\begin{fmtext}
\section{Introduction}\label{sec1}

The optical theorem, which relates the total scattering cross section of time harmonic  waves to the forward scattering amplitude, is a well known and useful  relation for acoustic, electromagnetic and elastic waves, e.g.\  \cite{Dassios00} derives the optical theorem   for both acoustics and elastodynamics.   Another  powerful general identity is for the {\it integrated extinction}, (IE), an integral of the scattering cross-section  over all wavelengths, which is a 
  useful measure of the total scattering strength of an acoustic target. 
The concept of integrated extinction was introduced in 1969 by Purcell \cite{Purcell69} in relation to optical scattering from interstellar grains.   The  IE is proportional to a linear combination of the monopole and dipole amplitudes if the scattering is {\it causal} \cite{Sohl07}, that is, the wavefront of the scattered response in the forward Fvers
direction arrives after an equivalent plane wavefront in the background medium.    Causal scattering is the default in electromagnetics since nothing travels faster than light in a vacuum; but there is no such limitation in acoustics.  Many scattering situations  of interest in acoustics are non-causal, such as metal objects in water or air, for which the IE expression of \cite{Sohl07} does not apply.

%\end{fmtext} %%%%%%%%%%%%%%% End of first page %%%%%%%%%%%%%%%%%%%%%

Our objective in this paper is an expression for the acoustic  integrated extinction that is valid  under all circumstances, one that is not restricted to causal or non-causal scatterers. 
Previous  forms of the causal  identity for  the IE have been cast in terms of the 
the frequency domain function %$S(\omega)$ 
that defines the forward scattering amplitude \cite{Sohl07}.   We find it more useful, particularly for  non-causal scattering, to phrase results in terms of the time dependent forward scattering function  %$s(t)$ 
which is the inverse Fourier transform of the frequency domain function. %$ S(\omega)$.  
\cite{Gustafsson2010} considered  EM forward scattering sum rules in the time domain, including the IE formula,   however, the results are restricted to causal scattering and therefore not  transferable to acoustics.   Note: the adjective "causal" is used here in the same sense as in \cite{Sohl07}.   Terms such as "post-scattering", "subsonic" could be used; however "causal" is preferred because  of its connection with analytic properties of complex-valued functions, which will be used later.  

The integrated extinction is a measure of total scattering strength over all frequencies which serves as a natural metric for measuring  reduction in  scattering using physical mechanisms \cite{Sohl2007,Monticone13}.  
The  general expressions for IE derived in this paper provide new interpretations for reduction of acoustic scattering. In contrast to the electromagnetic situation, it is possible to have the formal   IE expression for  causal acoustic scattering to become  zero. We show that this is possible iff the acoustic scatterer is non-causal. An important example of such an acoustic scatterer is the {\it neutral acoustic inclusion}, which  by definition has  zero monopole and dipole scattered amplitudes.   Based upon the IE results derived in this paper we  explore the relationships between integrated extinction, neutral acoustic inclusions, cloaking and causality.  

The outline of the paper is as follows.  In \S\ref{sec2} the scattering cross-section and integrated extinction are defined 
and the  result of \cite{Sohl07}, originally given   for 3-dimensional scattering is presented for causal acoustic scattering in one, two and three dimensions.  The main result for the IE is derived in \S\ref{sec3}.   Examples are given in \S\ref{sec4} for scattering in a one dimensional system for which the IE can be found in explicit form.

%Cloaking of  electromagnetic waves at a given frequency requires a superluminal phase velocity, which is physically permissible; however, as pointed out in \cite{Pendry06}, cloaking over a range of frequencies requires group velocities in excess of the speed of light, which is not possible.   This argument against the impossibility of EM invisibility over a bandwidth is based  on relativistic limits and dispersion.  No such limitations exist in acoustics. 

\section{Scattering cross section and integrated extinction}\label{sec2}

%\subsection{Scattering cross section}

Let $p({\bf x})$ be the  complex-valued acoustic pressure in a fluid of uniform density $\rho$ and compressibility $C$ (or its inverse the bulk modulus $K = C^{-1}$) outside of a finite region $\Omega$, the scatterer.  
It satisfies the Helmholtz equation
 \beq{1}\nabla^2 p + k^2 p = 0, \ \  {\bf x}\not\in \Omega , 
\eeq
where $k=\omega /c$, $\omega$ is frequency  and $c = 1/\sqrt{C\rho}$ is the sound speed. Time harmonic dependence is considered with the factor  $e^{-i\omega t}$ understood and omitted.   
 The system may be one, two or three-dimensional. 

The total  field is an incident plane wave %$p_{in} =p_0 e^{i k x}$ 
plus the   scattered pressure: $
p = p_0 e^{i k x} + p_s$  ($p_0$ constant).
Introduce the far-field scattering amplitude $S(\theta ,\omega)$ where $\theta$ defines the scattering direction, %$\hat {\bf x} =  {\bf x} /| {\bf x}|$ 
 $\theta = 0$ corresponding to the direction of incidence $\hat{\bf k}$, and 
\beq{3}
p_s ({\bf x}) =   p_0\, S(\theta ,\omega) \, 
%\Big( \frac{k}{i 2\pi | {\bf x}|}\Big)
\big[ k/(i 2\pi | {\bf x}|)\big]
^\frac{d-1}2 \, e^{ik | {\bf x}|}, 
\ \ | {\bf x}| \to \infty  %\ d = 1,2,\text{or}\ 3 .
\eeq
where $d=1$, $2$ or $3$ is the dimension.  
Equation \eqref{3} is exact in 1D $(d=1)$ for all ${\bf x}\not\in \Omega$ in which case $T= 1+S(0,\omega)$ and  $R=S(\pi ,\omega)$ are the transmission and reflection coefficients, respectively, satisfying $|R|^2 + |T|^2=1$. 
Define the total scattering cross section $\sigma$ (also known as the extinction) 
\beq{4}
\sigma (\omega)=  |p_0|^{-2}\,\int |p_s|^2 \dd s 
\eeq
where the integral is around a closed surface enclosing the scatterer. 
 Conservation of energy requires zero total energy flux across the closed surface, which 
%\beq{2}
%\int \big( |p_s|^2   + 2 \, \Re   p_s \bar p_{in} \big) \dd s = 0 
%\eeq
%where $\bar p_{in}$ is the complex conjugate. 
%Equations \eqref{3}, \eqref{4} and  \eqref{2} together 
implies the "optical theorem" relating  $\sigma$  to the forward scattering 
\beq{5}
 \sigma  = - 2\, \Re S(\omega)  \ \ \text{where} \  S(\omega) \equiv S(0,\omega) .
\eeq
Our  focus is on the  integrated extinction, defined as 
\bal{27}
\IE \equiv 
\int_0^\infty \frac {\sigma(\omega )}{\omega^2}\, \dd \omega  .
\eal
This is clearly a measure of the total scattering strength over all frequencies.  It is worth noting that $\sigma $ typically tends to a constant non-zero value as $\omega \to \infty$ while $\sigma ={\cal O}(\omega^2)$ for $\omega \to 0$ (although see the remark after eq\ \eqref{24}). 
Hence the only integer value $n$ for which the integral $\text{I}=\int_0^\infty \omega^{-n}\sigma \dd \omega$  is bounded, and therefore meaningful, is $n=2$, i.e.\  $\text{I}=\IE$.  The  IE  can also be expressed  as an integral over wavelength $\lambda = 2\pi c/\omega$, 
\bal{279}
\IE = \frac 1{2\pi c}
\int_0^\infty  \sigma \, \dd \lambda  .
\eal

%The forward scatter is a function of frequency, which is made clear through $S(0) \to S(\omega)$.   
Causality for the  forward scatter is defined such that wave motion  in the 
forward direction does not precede signals from the otherwise uniform medium.  This implies  that $S(\omega)$ is analytic in the upper half of the complex $\omega-$plane.   The transform $f(\omega )$ of any such causal function  satisfies the Sokhotski-Plemelj relations
for real values of $\omega$ \cite[eq.\ (1.6.11)]{Nuss72} 
\beq{6}
f(\omega )
 = \frac 1{i\pi } \dashint_{-\infty}^\infty 
\frac {f(\omega ')\dd \omega '}{\omega '- \omega }, 
\eeq
%\beq{6}
%\big( \Re f(\omega ),\, \Im f(\omega )   \big)  = \frac 1{\pi } \dashint_{-\infty}^\infty  \big( \Im f(\omega '),\, -\Re f(\omega ') \big)
%\frac {\dd \omega '}{\omega '- \omega }, 
%\eeq
where $ \dashint$ denotes  principal value integral. 
Setting $f(\omega ) =S(\omega )/\omega $, and using 
$\sigma(-\omega ) = \sigma(\omega )$, it follows that 
\beq{7}
\IE 
= \IE_\text{pc}
\ \ \text{where} \ \ 
\IE_\text{pc}  \equiv     \pi \Im \left. \frac{S(\omega )}{\omega}\right|_{\omega = 0} 
\eeq
and  "pc" is included to emphasize that this result is strictly limited to \emph{purely causal} forward scattering.  
Equation \eqref{7}, originally derived in  \cite{Sohl07} for 3D, is here generalized  to include the 1D and 2D cases. We note that eq.\ \eqref{7} also follows from the fact that $ -iS(\omega) $ is a 
Herglotz function\footnote{By definition, $h(z)$ is a Herglotz function if Im\,$h(z)\ge 0$ for Im\,$z>  0$.} \cite{Gustafsson10}. 

The zero frequency limit in \eqref{7} allows us to interpret $ \IE_\text{pc} $ in terms of quasistatic properties.  Thus, if the 
scatterer is a  region of volume $V$ with uniform density  $\rho '$ and compressibility  $C ' ({\bf x})$, then \cite{Sohl07} 
\beq{-2723}
\IE_\text{pc} 
= \frac {\pi}{2c} \Big( 
	\big( \frac{\langle C'\rangle}C -1 \big) V - \hat{\bf k}\cdot  
	{\boldsymbol \gamma}\Big(\frac{\rho}{\rho'} \Big)\cdot \hat{\bf k} \Big)
\eeq
where 
${\boldsymbol \gamma}$ is the polarizability dyadic \cite{Dassios00}  proportional to  $V$ and 
$\langle C' \rangle $ is  the average compressibility.

As noted in \cite{Sohl07}, the identity \eqref{7} is { only valid for causal scattering}.  Next, we show how to extend these results  to  non-causal scattering. 

\section{Integrated extinction for non-causal  scattering}\label{sec3}

Assuming the forward scattering is non-causal, we may  distinguish 
its causal and anti-causal components $s_+(t)$ and $s_-(t)$, respectively,  as      
\bal{-30}
s(t) \equiv \frac 1{2\pi} \int_{-\infty}^\infty S(\omega) e^{-i\omega t} \dd \omega  
= s_-(t)  + s_+(t)
 , \ \  s_\pm (t) = 0 \ \text{for}\ t  ^<_> 0.%\stackbin{<}{>} 0.
\eal
Thus, $s(t) = 0$ for $t<0$ if the scattering is causal, while for non-causal scattering there is some finite time before $t=0$ for which $s(t)\ne 0$.  
The time dependent function $s(t)$ can be identified as a forward scattered ``impulse-response'' function corresponding to an incident delta pulse $p_\text{inc} (x,t)=p_0 \delta (t - x/c)$.  
Subscripts $\pm$  denote functions that are causal/anti-causal, with Fourier transforms 
 analytic in the upper and lower half planes, respectively.  
%where $H(t) = 0,1$ for $t<0, t>0$ is the Heaviside function. 
The frequency dependent forward scattering function decomposes as 
\beq{-31}
S(\omega) 
= S_-( \omega)+S_+( \omega), 
\eeq
where
\beq{-313}
S_\pm( \omega) %= \pm  \int_0^{\pm \infty} s(t) e^{i\omega t} \dd t  
=\int_0^{\infty} s_\pm(\pm t) e^{\pm i\omega t} \dd t  .
\eeq
Causal scattering corresponds to  $s_-(t)=0$ $\Leftrightarrow$ $S_-( \omega)=0$; in the non-causal case both $s_-$ and $S_-$  are non-zero. 

We note the generalization of the Sokhotski-Plemelj relations \eqref{6},
\beq{63}
\big(
\Re f_\pm(\omega ),\, \Im f_\pm(\omega )
\big)
 = \pm \frac 1{\pi } \dashint_{-\infty}^\infty 
\big(
\Im f_\pm(\omega '),\, -\Re f_\pm(\omega ')
\big)
\frac {\dd \omega '}{\omega '- \omega } 
\eeq
for $\omega $ real.  Hence, using \eqref{-31}, %Using  $S( \omega)=S_+( \omega) +S_-( \omega)$ we have 
\bal{0-5}
-\frac 1{\pi }  \dashint_{-\infty}^\infty \Re S(\omega ) \frac {\dd \omega }{\omega^2} 
%&=-\frac 1{\pi }  \dashint_{-\infty}^\infty \Re S_+(\omega ) \frac {\dd \omega }{\omega^2} 
%- -\frac 1{\pi }  \dashint_{-\infty}^\infty \Re S_-(\omega ) \frac {\dd \omega }{\omega^2} 
%\notag \\
&= \left. \Im \frac{S_+(\omega )}{\omega}\right|_{\omega = 0+i0} 
-  \left. \Im \frac{S_-(\omega )}{\omega}\right|_{\omega = 0-i0} 
\notag \\
&= \left. \Im \frac{S( \omega )}{\omega}\right|_{\omega = 0} 
- \lim_{\epsilon \to 0} \Im 
  \bigg( \frac{S_-(i\epsilon )}{i\epsilon } 
+ \frac{S_-(-i\epsilon )}{-i\epsilon } \bigg) . 
\eal
This gives us the first form of our main result:
\beq{-32}\boxed{
\IE =
\pi \Im \bigg( \left. \frac{S(\omega )}{\omega}\right|_{\omega = 0} 
- 2  \frac{\dd S_-}{\dd \omega }(0) \bigg) .
}
\eeq
The second term involving $ S_-$vanishes if the scattering is causal, recovering the results of \cite{Sohl07}.
The   relationship \eqref{-32} can be written in alternative form   
\beq{-12}{
\IE =
\pi \Im \bigg(  \frac{\dd S_+}{\dd \omega }(0) -\frac{\dd S_-}{\dd \omega }(0) \bigg) 
}
\eeq
emphasizing the similar contributions from the causal and non-causal components of the forward scattered response.  
With  hindsight, the identity \eqref{-12} follows more directly by noting that  $\IE$  may be defined 
\bal{2721}
\IE = 
\int_0^\infty \frac {\dd \sigma }{\dd \omega }\, \frac{\dd \omega }{\omega} , 
\eal
which follows from \eqref{27} using the fact that $\lim_{\omega \to 0} \sigma / \omega =0$.

The identity \eqref{-12} does not have an obvious interpretation in terms of quasi-static quantities, unlike its causal version eqs.\ \eqref{7} and \eqref{-2723}.  However, if we represent  \eqref{-12} in the time domain, which is easily done using the definition of the Fourier transform, we find 
\beq{271}
\boxed{
\IE  =
\pi  \int_{-\infty}^\infty |t|\,  s(t)
   \dd t . }
\eeq
This result applies for causal and non-causal scattering.  Note the appearance of $|t|$ in \eqref{271}.  The same integral with  $|t|$ replaced by $t$ is simply  
\beq{2710}
  \int_{-\infty}^\infty t\,  s(t)
   \dd t 
	= \Im \left. \frac{S(\omega )}{\omega}\right|_{\omega = 0}.
\eeq
Hence, eqs.\ \eqref{7}, \eqref{271} and \eqref{2710} imply that the integrated extinction can be written  
\beq{-273}
\boxed{
\IE 
= \IE_\text{pc} 
	-2 \pi  \int_{-\infty}^0 t\,  s(t)
   \dd t  . 
	}
\eeq
This particular form of the integrated extinction identity generalizes the causal-only result \eqref{7}. 
Note that  the non-causal response is zero for some $t_- <0$, i.e. $s(t)=0$, $t<t_-$, and hence the integral in \eqref{-273} is only over a finite interval of time from $t_- $ to $0$.

\section{Examples} \label{sec4}

\subsection{Arbitrarily layered one dimensional medium}\label{sec4a}

Consider a 1D system with non-uniform density and compressibility  $ \rho '(x)$, $C'(x)$  restricted to $\Omega = [0,a]$.  The forward scattering amplitude is 
\beq{9}
S(\omega ) %=S(0, \omega ) 
= -1 + 2e^{-i k a} /\big( 
M_{11}+M_{22}- zM_{12}-z^{-1}M_{21}\big)
\eeq
where 
$z= \rho c$ is the background acoustic  impedance and $M_{ij}$ are elements of the 2$\times$2 propagator matrix $ {\bf M}(a)$, $\det {\bf M} = 1$, the solution to 
\beq{8}
\frac {\dd{\bf M}(x) }{\dd x}  = i\omega {\bf Q}{\bf M}, \ \ 
{\bf M}(0) = {\bf I}, 
\eeq
with ${\bf I}$  the identity and 
\beq{-56}
 {\bf Q}(x) = 
\begin{pmatrix} 0 & C'(x)  \\ \rho ' (x)& 0 
\end{pmatrix} .
\eeq
The solution \eqref{9} follows using standard methods for uni-dimensional systems.  For our purposes, we note that 
 at low frequency  ${\bf M}(a) = {\bf I} + i\omega a \langle {\bf Q} \rangle + ...$ where $\langle \cdot \rangle $ denotes the average value in  $\Omega$.  

Define   the non-dimensional travel time %through the heterogeneity, by 
\beq{698}
 \T  = \frac ca \int_0^a \frac {\dd x }{c'(x)}  
\eeq
which is    the ratio of the travel time across the inhomogeneous region to  the 
travel time across an equivalent uniform slab.  The  slab has the same travel time as the equivalent uniform slab if 
  $\T = 1$; hence the scatterer is causal if $\T \ge 1$, and non-causal otherwise.  Next, we consider the causal and 
non-causal cases separately. 

For causal scattering ($\T \ge 1$)  the integrated extinction is simply, using eq.\ \eqref{7},
\beq{10}
\IE_\text{pc} 
= \frac{\pi}2 \frac ac 
\Big( 
 \frac{ \big\langle C'  \rangle}C +\frac{\langle \rho' \rangle}\rho - 2 \Big).
\eeq
It is then a straightforward calculation to show that  
\bal{12}
\IE_\text{pc} 
= {\pi} \frac ac \bigg( \frac 12
\Big\langle  \frac c{c'}
\Big( \sqrt{\frac{z'}z} - \sqrt{\frac z{z'}}  \Big)^2 
%\Big( \big( \frac{\rho'}\rho \big)^{1/2} - \big( \frac{C'}C \big)^{1/2} \Big)^2 
\Big\rangle
+   \T  - 1  \bigg)  
\eal
where $z' = \rho ' c' $ is the  impedance. 
Since the scatterer is causal  $(\T \ge 1)$ this means that $\IE_\text{pc}  \ge  0 $ with equality iff $\T = 1$ and the local  impedance  is constant and equal  to the uniform impedance, i.e. $z' = z$.   In summary, it is possible that the total  IE  of eq.\ \eqref{10} can vanish, but only if the slab has (i) uniform impedance and  (ii) travel time equivalent to the uniform medium.   Condition (i) is expected, it guarantees no reflection/back-scattering, while (ii) ensures that the phase of the transmitted wave is the same as that for the uniform medium. 

Note that the  expression in the right hand side of \eqref{12} can vanish for a non-constant impedance if  
\beq{15}
\T = 1 - \frac 12 
\big\langle  \Big( 
\big( \frac{\rho'}\rho \big)^{1/2} - \big( \frac{C'}C \big)^{1/2}
\Big)^2 \big\rangle
\le 1 . 
\eeq
However,  \eqref{12}  no longer represents the IE  in this case since the scattering is 
 non-causal $(\T < 1 )$ and therefore $\IE\ne \IE_\text{pc} $.   That is, satisfaction of \eqref{15} does {\it not} imply that the integrated extinction vanishes.

An expression for the IE under both causal and non-causal conditions can be found if the impedance is everywhere constant. The transmitted wave is then of unit amplitude  with phase defined by $\T$ of \eqref{698}, and therefore $S(\omega ) = -1 + e^{-ika} e^{ika \T}$.  The split functions are   
$S_+(\omega ) =S(\omega )$, $S_-(\omega ) =0$ if $\T \ge 1$, and $S_+(\omega ) =0$, $S_-(\omega ) =S(\omega)$ if $\T <1$, from which the IE follows using eq.\ \eqref{-12} as 
\bal{121}
\IE 
= {\pi} \frac ac \, |\T  - 1|  \ \ \text{for constant impedance}.
\eal
%The analytical expression \eqref{12}  for the  IE does not apply for non-causal scattering, one must 

\subsection{A uniform layer}
A layer with material properties $\rho '$, $C'$ occupies $x\in[0,a]$ in the otherwise uniform background.  Then, setting $k ' = \omega /c'$, $c' = 1/\sqrt{\rho 'C'}$, $z'=\rho ' c'$, 
\bal{-15}
S(\omega ) = -1 + 4  z{z'} e^{-i ka} \big(
({z'} +  z)^2 e^{-i k'a} - ({z'} - z)^2 e^{i k'a} 
%S(\omega ) = -1 + 4 \frac z{z'} e^{-i ka} \Big[
%\big(1 + \frac z{z'}\big)^2 e^{-i k'a} - \big(1 - \frac z{z'}\big)^2 e^{i k'a} 
\big)^{-1} 
\eal
from which the forward scattering  impulse response is found 
\beq{-20}
s(t) = -\delta(t)  +   \frac {4zz'} {(z' - z)^2} \sum_{n=1}^\infty
\big(\frac{z'-z}{z'+z} \big)^{2n} \delta(t-t_n) , 
\  \   t_n = (2n-1)\frac{a}{c'}  - \frac ac .  
%, \ \ R = \frac{z'-z}{z'+z} 
\eeq
If the layer is non-causal, let $N\ge 1$ be the number of non-causal pulses, i.e.  $t_N=$max$\, t_n$, $t_n<0$, then eqs.\ \eqref{-31} and \eqref{-20} yield  
\bal{-16}
S_-(\omega ) = \frac {4zz'} {(z' - z)^2}  \sum_{n=1}^N \big(\frac{z'-z}{z'+z} \big)^{2n}
\, e^{i\omega t_n}. 
\eal
This may be summed to give 
\bal{-163}
S_-(\omega ) = \big(  S(\omega ) +1\big) \, \Big(1 - 
\Big[\Big(\frac{z'-z}{z'+z} \Big) e^{i\omega   \frac a{c'}}  \Big]^{2N}
\Big)  .
\eal

We now give three different expressions for the integrated extinction based on the equivalent identities \eqref{-32}, \eqref{271} and \eqref{-273}.  The first formula for $\IE$ corresponds to  eqs.\  \eqref{271} and it 
follows directly using  eq.\ \eqref{-20}, 
\beq{-19}
\IE 
= \frac {4\pi zz'} {(z' - z)^2} \sum_{n=1}^\infty \big(\frac{z'-z}{z'+z} \big)^{2n} \, |t_n|.
\eeq
The strictly causal expression for $\IE $ is given by eq.\ \eqref{10}, and the non-causal variant follows from the  part of $s(t)$ of \eqref{-20} that is non-zero for negative time, i.e. for $t_n<0$.  Combining these yields an alternative expression for  
$\IE$ corresponding to  eq.\  \eqref{-273}, 
\beq{-17}
\IE 
= \frac{\pi}2 \bigg( \Big( \frac{z}{z'} +\frac{z'}{z} \Big) \frac a{c'}   -2\frac ac 
- 16 \frac z{z'} \big(1 - \frac z{z'}\big)^{-2} \sum_{n=1}^N \Big(\frac{z'-z}{z'+z} \Big)^{2n} \, t_n
\, H(-t_1)
\bigg)   
\eeq
where $H$ is the Heaviside step function.  Finally, using eq.\ \eqref{-163}, the formula  \eqref{-32} becomes for this example
\beq{2-17}
\IE 
= \frac{\pi}2 \Big(  \Big( \frac{z}{z'} +\frac{z'}{z} +4N\Big) \frac a{c'}   -2\frac ac   \Big) 
\Big( 2\Big(\frac{z'-z}{z'+z} \Big)^{2N} - 1 \Big)
+  2\pi N\frac a{c'}  .
%\frac{\pi}2 \Big(  \frac a{c'} \Big( \frac{z}{z'} +\frac{z'}{z} \Big)   -2\frac ac   \Big) \Big( 2\Big(\frac{z'-z}{z'+z} \Big)^{2N} - 1 \Big) + N 4\pi \frac a{c'}  \Big(\frac{z'-z}{z'+z} \Big)^{2N} .
\eeq

The  three   expressions \eqref{-19},  \eqref{-17} and  \eqref{2-17} for  $\IE $ are obviously equivalent, and are each  valid whether the scattering is causal or non-causal. Equation \eqref{-19} gives the integrated extinction as an infinite sum, eq.\ \eqref{-17} replaces the infinite sum with  a finite one, and eq.\ \eqref{2-17} provides the most compact version.  Note that the simpler result for causal scattering, i.e.\ eq.\ \eqref{10}, is evident from   eqs.\ \eqref{-17} and  \eqref{2-17}, corresponding to $N=0$ in the latter. The parameter $N$ is a positive integer for non-causal scattering.  This might suggest that the expressions \eqref{-17} and  \eqref{2-17} behave discontinuously as a function of the slab speed $c'$, which defines  $N$.  However, this is not the case as can be seen from the fact that as $N$ increases or decreases by unity, one additional  value of $t_n$ is zero as it becomes negative or positive, respectively.  The identities 
\eqref{-19} and   \eqref{-17} are then clearly continuous functions of $c'$, as must be eq.\ \eqref{2-17}. 
%%%%%%%%%%%%%%%%%%%%%%%%%%%%%%%%%%%%%%%%%%%%%%%%%% Figure
\begin{figure}[H]
				\begin{center}	
				\includegraphics[width=4.6in 	]{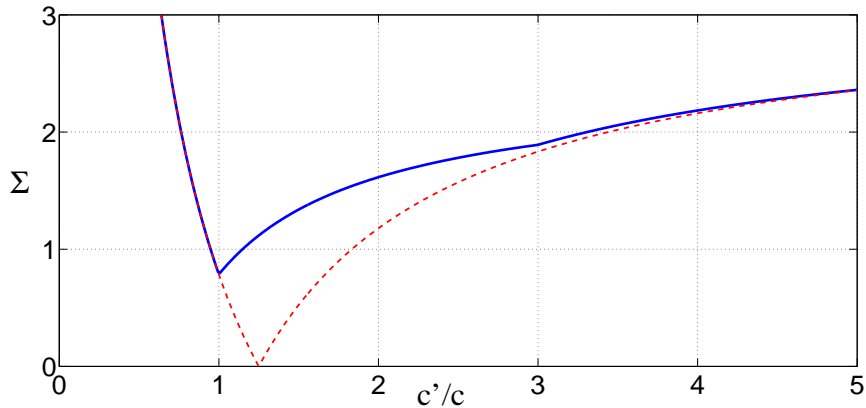} 
	\caption{ The solid curve shows $\IE$  as a function of $c' /c$ for a uniform slab $z'/z=2$, $a/c=1$.  The dashed curve shows the magnitude of the causal integrated extinction  defined in \eqref{3-4}. }
		\label{fig1} \end{center}  
	\end{figure}
%%%%%%%%%%%%%%%%%%%%%%%%%%%%%%%%%%%%%%%%%%%%%%%%%% 

\subsection{Numerical example}

The integrated extinction for a uniform slab with impedance twice the background value $(z'  = 2z)$ is shown in Figure \ref{fig1} as a function of the slab wave speed.  Also shown is the \emph{magnitude} of the purely causal  IE  (see  eq.\ \eqref{10}) 
\beq{3-4}
\IE_\text{pc} = 
\frac{\pi a} {2c} \Big(  \Big( \frac{z}{z'} +\frac{z'}{z} \Big) \frac c{c'}   -2    \Big)  .
\eeq
Clearly, $\IE  = \IE_\text{pc}  $ for $c' \le c$ as expected.  However, it is also evident that $\IE  \approx - \IE_\text{pc}  $ for  $c' >3c$ with the approximation better as $c'$ becomes  larger.   The reason for this can be understood from the mathematical form of  $\IE $ in eq.\ \eqref{2-17}.  The number $N$ increases approximately linearly with the slab speed $c'$when  all other slab parameters are held fixed.  The quantity $(\frac{z'-z}{z'+z} )^{2N}$ therefore decreases with increasing $c'$, so that $\IE \to -\IE_\text{pc}  $ as this quantity tends to zero.

%%%%%%%%%%%%%%%%%%%%%%%%%%%%%%%%%%%%%%%%%%%%%%%%%% Figure
\begin{figure}[H]
				\begin{center}	
				\includegraphics[width=4.6in ]{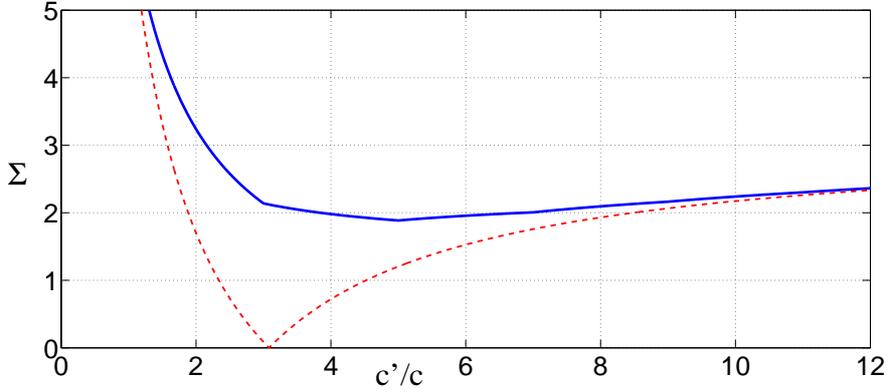} 
	\caption{ The curves are the same as in Figure \ref{fig1} except that the slab impedance is $z'=6 z$. }
		\label{fig2} \end{center}  
	\end{figure}
%%%%%%%%%%%%%%%%%%%%%%%%%%%%%%%%%%%%%%%%%%%%%%%%%% 
Figure \ref{fig2} shows the same phenomenon for a larger slab impedance $(z'/z = 6)$.  The limiting behavior $\IE  \approx - \IE_\text{pc}   $ is again observed, this time at larger values of $c'$ since the quantity $(\frac{z'-z}{z'+z} )^{2N}$ decreases less rapidly than for Figure \ref{fig1}.  
Note that the same curves are obtained in Figures \ref{fig1} and \ref{fig2} for $z/z'=2$ and 
$z/z'=6$, respectively,  since eq. \eqref{2-17} is unchanged for $\frac {z'}z \leftrightarrow \frac z{z'}$. 

%%%%%%%%%%%%%%%%%%%%%%%%%%%%%%%%%%%%%%%%%%%%%%%%%% Figure
\begin{figure}[H]
				\begin{center}	
				\includegraphics[width=4.2in 	]{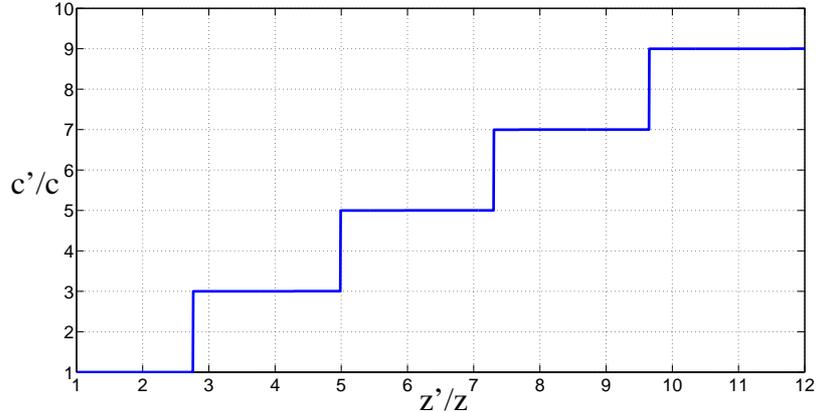} 
	\caption{ The curve shows the value of slab wave speed $c'$ which minimizes the integrated extinction of eq.\ \eqref{2-17} for  given values of slab impedance  $z' \ge z$, and fixed slab thickness.    }
		\label{fig3} \end{center}  
	\end{figure}
\begin{figure}[h]
				\begin{center}	
				\includegraphics[width=4.2in 	]{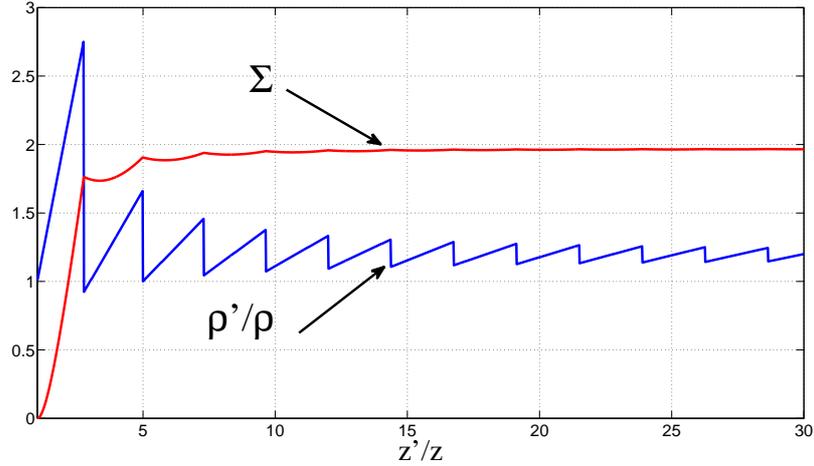} 
	\caption{ For given values of slab impedance  $z'\ge z$ and $a/c=1$, the blue curve shows the value of slab density $\rho '$ which minimizes the integrated extinction of eq.\ \eqref{2-17}.  The red curve is the corresponding minimal value of  $\IE$.  }
		\label{fig4} \end{center}  
	\end{figure}

%%%%%%%%%%%%%%%%%%%%%%%%%%%%%%%%%%%%%%%%%%%%%%%%%% 
The simple form of the expression for the  purely causal IE in  eq.\ \eqref{3-4} indicates that for a given value of the slab impedance, $z'$,  the minimum value of $\IE_\text{pc}   $ for 
all possible slab wave speeds $c' \le c$ is achieved at $c'=c$.  Conversely,   for a given value of $c' <c$,  the minimum value of $\IE_\text{pc}   $  is achieved at $z'=z$.        
The location of the minimizing values of the dual parameter at the edges $c'=c$, $z'=z$ suggest that these   IE values are not necessarily the global minima.   Figure \ref{fig3} shows, as expected,  that the global minimum of IE for a given slab impedance is achieved by a non-causal slab wave speed.  The minimizing value of  $c'/c$ is an odd integer greater than or equal to unity, the value of which is a  discontinuous function of $z'/z$ that increases monotonically in a roughly  linear manner.  The complicated nature of this function  is displayed in Figure \ref{fig4} which shows the 
minimizing value of the slab density as a function of slab impedance. 

%%%%%%%%%%%%%%%%%%%%%%%%%%%%%%%%%%%%%%%%%%%%%%%%%% Figure
%%%%%%%%%%%%%%%%%%%%%%%%%%%%%%%%%%%%%%%%%%%%%%%%%% 

\subsection{Two and three dimensions}%\label{sec4d}

Consider plane wave incidence on a uniform circular $(d=2)$ or spherical $(d=3)$ inhomogeneity of radius $a$ and  properties $C'$, $\rho '$.  %  in the background acoustic medium $C$, $\rho$.
The low frequency expansion of the forward scattering amplitude follows, using standard methods, as
\beq{23}
S( \omega) = i\omega \pi \frac{(d-1)}d \,  \frac{a^d}c \Big[ \frac{C'}C - 1 
+ \frac{d\big(\frac{\rho'}\rho  - 1\big) }{(d-1)\frac{\rho'}\rho  + 1}  
\Big] 
 + {\cal O}(\omega^2). %\ldots .
\eeq
The IE for causal scattering then follows from eqs.\ \eqref{7} and \eqref{23} as  
\beq{24}
\IE_\text{pc}
=  \pi^2 \frac{(d-1)}d \, \frac{a^d }c\, \frac \rho{\rho'}
\bigg(
\frac{\big(\frac{\rho'}\rho  - 1\big)^2 }{(d-1)\frac{\rho'}\rho  + 1}  
+  \T^2 - 1 
\bigg)
\eeq 
where $\T = (\rho 'C')^{1/2} /c$ is the ratio of travel times.  
It is interesting to compare this with the analogous 1D result  eq. \eqref{12}.   Since $\T \ge 1$ for causal scattering, eq.\ \eqref{24} shows that $\IE_\text{pc}$ vanishes iff $\rho '=\rho$, $C' =C$, 
implying that there is no combination of $\rho '$, $C'$ that is  transparent except for the trivial case when the scatterer is identical with the background acoustic fluid. 

The causal IE of eq.\ \eqref{24} tends to infinity as the compressibility $C'$ of the scatterer becomes infinite. %    
This limit  can be identified with pressure release  boundary conditions:  $p=0$ on $r=a$, for which the scattering cross-section of the sphere tends to a non-zero constant as $\omega \to 0$, specifically $\sigma = 4\pi a^2 \big[ 1 + \frac 13 (ka)^2 + {\cal O}\big( (ka)^4\big)   \big]$     \cite{Athanasiadis2001}.   The IE as defined in eq.\ \eqref{27} is not integrable at $\omega =0$ and therefore not applicable to scatterers with  pressure-release boundaries.  However, such boundary conditions are not strictly physical since a fluid with a pressure-release inclusion is not stable under static pressure perturbation: the hole collapses.  This non-physical limit is apparent in the finite  scattering cross-section of the sphere at zero frequency.

\section{Discussion}\label{sec5}

\subsection{The limiting cases $\IE= \IE_\text{pc} $ and $\IE \approx -\IE_\text{pc} $}

The above results suggest   separation of   
the integrated extinction as distinct contributions
\beq{3-12} 
\begin{aligned}
\IE^{(+)}  &=
\pi  \int_{0}^\infty t\,  s(t)
   \dd t  =
\pi \Im    \frac{\dd S_+}{\dd \omega }(0)  ,
\\
\IE^{(-)}  &= -
\pi  \int_{-\infty}^0 t\,  s(t)
   \dd t  = - 
\pi \Im    \frac{\dd S_-}{\dd \omega }(0)  .
\end{aligned}
\eeq
$\IE^{(+)} $ and $\IE^{(-)} $ can be considered as  causal and non-causal  parts of the  IE, respectively, through the relation of the IE to the forward scattering.  Their connection with the general and the purely causal expressions for the   IE  follow from   eqs.\ \eqref{-12}, \eqref{271} and    \eqref{-273}, 
\beq{3-22} 
\begin{aligned} \IE  &=
  \IE^{(+)} +\IE^{(-)} ,
\\  \IE_\text{pc}  &= 
\IE^{(+)} -\IE^{(-)} 
.
\end{aligned}
\eeq

The IE can be expressed in the suggestive forms
\beq{-27} 
\begin{aligned} \IE  &= \IE_\text{pc}  + 2  \IE^{(-)} 
\\   &= -\IE_\text{pc}  + 2\IE^{(+)} 
.
\end{aligned}
\eeq
 The identity  of Sohl et al.\ \cite{Sohl07} in eq.\ \eqref{7} for purely causal scattering   follows from the first of these  with $\IE^{(-)} =0$.  
A major point of this paper is that  the scattering can be \emph{mainly} non-causal in the sense of the examples in Figures \ref{fig1} and \ref{fig2}.  This can occur in one, two or three dimensions if the scatterer supports very fast waves relative to the background,  e.g.\   metallic scatterers in air.  If the non-causal scattering predominates then most of the incident energy is scattered before $t=0$, implying  $|\IE^{(-)} | \gg  |\IE^{(+)}| $ and therefore 
\beq{-23}
\IE 
\approx -\IE_\text{pc}  .
\eeq
Relation \eqref{-23} is essentially the opposite of the causal identity \eqref{7}.   Introduce a new quantity, $\IE_\text{pnc} \equiv -\IE_\text{pc}$, then it is clear that $\IE_\text{pnc}$ is the IE for purely non-causal scattering. 
 It should be borne in mind that   the approximation \eqref{-23} becomes precise only in the limit that the scatterer supports waves of infinite speed, which is not possible in acoustics because it requires either zero material density or compressibility.  Of course, infinite wave speed is prohibited in electromagnetics, although for a different reason. 

%The physical explanation for the fact that the total integrated energy scattered is the negative of the causal IE in eq.\ \eqref{-23} can be understood mathematically from eq.\ \eqref{3-22}. 

\subsection{Neutral acoustic inclusions and cloaking}

For purely causal scattering, as in electromagnetics, the IE must be positive definite and hence  the forward scattering amplitude on the right hand side of \eqref{-2723} must be positive.  This constraint does not apply in acoustics. 
The multipole expansion of the scattering function $S(\theta,\omega)$ at low frequency (long wavelength) is a power series in $\omega$, hence the low frequency limit of the forward scatter is necessarily defined by the first few multipole terms.   
%--------
A {\it neutral acoustic inclusion} by definition has  zero monopole and dipole scattered amplitudes. 
The two terms in the expression \eqref{-2723} for $\IE_\text{pc}$  correspond to the monopole and dipole contribution to the forward scatter.  Specifically, the monopole amplitude is proportional to 
$\big(\frac{\langle C'\rangle}C -1 \big) V $ while the dipole amplitude depends on the    polarizability tensor 
$	{\boldsymbol \gamma}$.     Both terms are identically zero for a neutral acoustic inclusion, implying that $\IE_\text{pc} =0$.  

Hence,  \emph{ a neutral acoustic inclusion is a  non-causal scatterer}.  By definition 
\beq{272}
  \int_{-\infty}^\infty t\,  s(t)
   \dd t =0 \ \ \text{for a neutral acoustic inclusion},
\eeq
which follows from eqs.\ \eqref{7} and \eqref{2710}. 
Combining eqs.\ \eqref{271} and  \eqref{272} yields
\beq{273}
\IE 
= -2 \pi  \int_{-\infty}^0 t\,  s(t)
   \dd t \ \ \text{for a neutral acoustic inclusion}. 
\eeq
The integrated extinction of a neutral acoustic inclusion is therefore defined completely by the non-causal part of the forward scatter. 

The concept of neutral inclusions is quasistatic: from this point of view a neutral inclusion  inserted in a matrix containing a uniform applied field does not disturb the field outside \cite[pp.\ 113-142]{Milton01}. Arrays of neutral inclusions therefore have effective static moduli that can be determined exactly. Coated spheres or cylinders provide the simplest examples of  neutral inclusions.  
%The problem of finding neutral inclusions goes back at least to the 1953 paper by the work of Mansfield \cite{Mansfield53}, who found that certain reinforced holes could be cut out of a uniformly stressed plate without disturbing the surrounding stress in the plate.  
The acoustic equivalent of the quasistatic definition   is that the monopole and dipole amplitudes vanish in the long wavelength limit for  plane wave incidence in any direction.  
The vanishing of the monopole and dipole amplitudes leads to reduced scattering in the low frequency range, resulting in increased transparency.   This connection with invisibility was noted as early as 1975 by Kerker \cite{Kerker75}. 
Neutral inclusions have been proposed as models for  transparency of 
 EM waves by \cite{Alu05} with many subsequent developments.  Thus, \cite{Zhou07} 
 examined   acoustic transparency for a multilayered sphere;  \cite{Liu10a} consider applications in electromagnetic shielding. The analogous elastic problem was considered by \cite{Zhou08}.   Cylindrical elastic shells in water \cite{Titovich14a} provide a practical design for neutral acoustic inclusions: the shell thickness can be adjusted to give the appropriate effective compliance, $\langle C'\rangle = C$, while simultaneously matching the effective density so that $	{\boldsymbol \gamma} =0$.    The shell then has reduced total-scattering cross section in the long-wavelength regime \cite{Titovich14a}.    

The neutral acoustic inclusion conditions   reduce to the following constraints on the  compressibility and density of the scatterer:
 \beq{5-5}
 \langle C'  \rangle =C , \ \  
\langle \rho' \rangle = \rho  .
\eeq
These conditions guarantee that the forward scattering quantity $\IE_\text{pc}$ vanishes for every direction of incidence, leading to reduced low frequency scattering, or increased transparency.    
The notion of causality, or relative signal speed, is not contained in the quasistatic definition of neutral inclusions. The above finding that  neutral acoustic inclusions must be   non-causal scatterers is therefore remarkable in that a  quasistatic effect implies a necessary dynamic property.      

Transparency over a wide range of frequencies, known as {cloaking},  demands  at the very least that the object be transparent in the long wavelength regime, which may be achieved by making the cloak and the object being cloaked together satisfy \eqref{5-5}.    We  therefore conclude that \emph{low frequency acoustic cloaking requires non-causal scattering}.   However, if the object+cloak  is a non-causal scatterer then the first arrival, i.e.\  the high frequency forward scatter, is non-zero and hence the IE is non-zero and the cloaking is imperfect.  Conversely, if the forward scatter first arrival is at $t=0$ then the IE integral \eqref{273} is zero and the cloaking is perfect.  This means that in order to achieve perfect cloaking one \emph{only} needs (i) to satisfy the low frequency cloaking conditions \eqref{5-5}, and (ii) have the forward signal arrive at $t=0$ for all  directions of incidence. The adverb "only" is in italics to emphasize that it is not clear whether or not these conditions can be simultaneously satisfied  in 2D and 3D except in the trivial case $C'=C$, $\rho'=\rho$.  By comparison, $\IE =0$ is possible for non-trivial $C'$, $\rho'$ in 1D, as shown in eq.\ \eqref{121}. 
The condition (ii) is similar to the "eikonal" condition, a weak form of transformation acoustics; the latter is a pointwise mapping of the  wave equation while the former only maps rays \cite{Norris08b}.  From a physical point of view, as compared with a purely mathematical one, it  can be argued that condition (ii) is unattainable because the cloaking material must contain microstructure which sets an upper limit on frequency.  

%Therefore, cloaking cannot be simultaneously achieved at the low and high frequency limits, i.e.\ quasistatic cloaking and ray-based transparency are mutually incompatible.  More generally, acoustic cloaking must be band limited.   It is interesting to compare this with the   stronger   conclusion for electromagnetic waves \cite{Pendry06} which is based on the impossibility of faster-then-light signals, i.e.\  that the cloak must be dispersive and  therefore only effective at a single frequency. }

Despite the impossibility of acoustic invisibility at all frequencies, the present results for IE provide a new means of characterizing broadband transparency.  Let us  identify IE as a metric of acoustic transparency, since perfect transparency/invisibility corresponds to $\IE=0$.   A first step towards achieving broadband transparency is to convert the target into a neutral acoustic inclusion, by surrounding it with a ``low frequency cloak'' or otherwise.  Further improvement in broadband transparency is then equivalent  to minimizing the integral \eqref{273}.   The significance is that \eqref{273} is an  integral over a finite time span, from the (negative) time of the first arrival in the forward direction until $t=0$.  This  provides a  new time-domain based metric for broadband transparency that depends only on the part of the signal that arrives before the background  wave.   The benefit of this approach is that it is restricted to a finite length time domain signal, while it provides a scalar measure of the complete broadband scattering properties.  The challenge is to understand the dependence of the IE of eq.\ \eqref{273} on  design parameters, such as cloak properties.

\section{Conclusion}   \label{sec6}

We  have derived a relation  between the integrated extinction and the forward response that is valid for non-causal scatterers, generalizing the known strictly causal identity \cite{Sohl07}.   The result can be expressed in terms of  frequency dependent functions, eqs.\ \eqref{-32} and \eqref{-12}, or using time dependent forward scattering functions, eqs.\ \eqref{271} and \eqref{-273}.    The time dependent representation, which is quite different from previous frequency-based formulae for the IE, leads to interesting implications for the acoustic transparency and cloaking.   The connection is through the IE expression for purely causal scattering, $\IE_\text{pc} $ of eq.\ \eqref{-2723}.  This quantity can vanish for a wide variety of scatterers, including those we call neutral acoustic inclusions.  These are acoustically transparent in the long wavelength limit, an example of low frequency cloaking. We have shown here that  neutral acoustic inclusions are necessarily non-causal scatterers with IE given by the finite integral \eqref{273}.  This expression for the IE involves only the part of the forward scattered signal arriving before $t=0$ and as such is a new metric for  improving cloaking starting with  an effective low frequency neutral acoustic inclusion.   One possible approach would be to vary cloaking parameters to minimize the integral \eqref{273} and thereby improve the bandwidth of the low frequency cloak.  

The examples of \S\ref{sec4} for  1D systems show that the  IE can be found in explicit form for causal scattering from an arbitrary region of inhomogeneity, eq.\ \eqref{12}.  Three distinct but equivalent forms of the IE are given in eqs.\ 
\eqref{-19} to \eqref{2-17} for scattering from a uniform region of inhomogeneity, causal or non-causal.   This is the first time that an expression for the IE has been presented for non-causal scattering.    Future studies will consider 2D and 3D examples.

\section*{Acknowledgment}
%\ack
{This work  was supported under ONR MURI Grant No. N000141310631}

%%%%%%%%%%%%%%%%%%%%%%%%%%%%%%%%%%%%%%%%%%%%%%%%%%%%%%%%%%%%%%%%%%%%%%%%%%
%\bibliography{../../SHARED_BIBLIOGRAPHY/AN_BIG_BIB}
%\bibliographystyle{bibgen}%unsrt}%natbib}%unsrtnat}%doipubmed}%harvard}% plain}%uabbrvnat}%
%\end{document}

\end{document}